\documentclass[%
reprint,
footinbib,
amsmath,amssymb,
]{revtex4-1}

\usepackage{graphicx}
\usepackage{dcolumn}
\usepackage{bm}
\usepackage{color}

\begin{document}


\title{Large systems of random linear equations with non-negative solutions: \\ Characterizing the solvable and unsolvable phase}

\author{Stefan Landmann }
 \altaffiliation[Correspondence]{}
  \email{stefan.landmann@uni-oldenburg.de}
\author{Andreas Engel}%

\affiliation{%
Institute of Physics, Carl von Ossietzky University of Oldenburg,
D-26111 Oldenburg,
Germany
}%


\date{\today}

\begin{abstract}
Large systems of linear equations are ubiquitous in science. Quite often, e.g. when considering population dynamics or chemical networks, the solutions must be non-negative. Recently, it has been shown that large systems of random linear equations exhibit a sharp transition from a phase, where a non-negative solution exists with probability one, to one where typically no such solution may be found. The critical line separating the two phases was determined by combining Farkas' lemma with the replica method. Here, we show that the same methods remain viable to characterize the two phases away from criticality. To this end we analytically determine the residual norm of the system in the unsolvable phase and a suitable measure of robustness of solutions in the solvable one. Our results are in very good agreement with numerical simulations. 
\end{abstract}

\pacs{Valid PACS appear here}
\maketitle


\section{\label{sec:Introduction}Introduction}
Systems of linear equations are fundamental objects of study in linear algebra. They play an important role in many fields, ranging from physics to ecology and financial market analysis. In many situations, e.g. when examining the stability of stationary states of systems with a large number of degrees of freedom, one encounters systems with many equations. 

In the study of large complex systems it is often not known in detail how the microscopic parts interact with each other. Fortunately, the macroscopic properties frequently do not depend on all of the microscopic details, and it has proven successful to model them by random variables. This is a sensible approach if so-called self-averaging quantities exist that depend only on the parameters of the distributions and not on the individual realizations. A classic example is spin-glass theory~\cite{edwards1975theory,sherrington1975solvable}, but also problems from computational complexity~\cite{monasson1999determining}, information theory~\cite{mezard2009information}, and artificial neural networks~\cite{engel2001statistical} have been analyzed along these lines.

A natural question occurring when studying large systems of random linear equations concerns their solvability. In some cases this question can be answered easily. Consider a system of linear equations $\hat{a}^T \mathbf{x}=\mathbf{b}$ for the variables $x_\mu \in \mathbb{R}, \mu=1,\dots, S$ with a real $N \times S$  matrix $\hat{a}^T$ and a real inhomogeneity vector $\mathbf{b}\in \mathbb{R}^N$. In general, this system has a solution if the rank of the matrix, $r(\hat{a}^T)$, is the same as the rank of the augmented matrix, $r(\hat{a}^T | \mathbf{b})$. If the entries of the matrix are drawn independently from a continuous probability distribution it has full rank with probability one \cite{feng2007rank} and the system is almost surely solvable for $S \geq N$, i.e. if there are at least as many variables as equations.

In many situations, e.g. when considering models describing population dynamics~\cite{may1972will,Schuster,tikhonov2017collective}, chemical networks~\cite{PoEs,Schnakenberg}, financial markets~\cite{kondor2017analytic}, or game-theoretic settings~\cite{berg1998matrix} the searched-for variables are concentrations or probabilities and must therefore be non-negative. The question whether a solution $\mathbf{x}$ of the system of linear equations exists with all components $x_\mu$ {\em non-negative} is non-trivial if $S \geq N$.

In~\cite{landmann2019non} it was shown that this problem may be mapped onto a dual one by using Farkas' lemma~\cite{farkas1902theorie}. For independent random entries $a_{\mu i}$ and $b_i$ and $S=\mathcal{O}(N)$ the dual problem is amenable to a replica analysis. Remarkably, in the limit $N,S\to \infty$ the system is characterized by a sharp transition from a phase in which a non-negative solution exists with probability one, to one where typically no such solution can be found. The transition line only depends on the statistical properties of the system and is therefore self-averaging.

In~\cite{landmann2019non} the mapping to the dual problem was used solely to determine the critical line between the two phases. In the present paper we show that it remains valuable also to characterize the system {\em away} from criticality, i.e. deeply inside the solvable and unsolvable phase, respectively. In this way we get analytic expressions for the remaining variability of the system in the solvable phase as well as for the residual error in the unsolvable one. 

The paper is organized as follows. In Section~\ref{Sec: Problem and notation} the problem is defined, relevant notation is fixed, and suitable quantities to characterize the two different phases are introduced. Section~\ref{Sec: Farkas} contains an intuitive explanation of Farkas' lemma, which is central to this paper. In Section~\ref{Sec: Determination of critical line} we sketch the determination of the critical line as performed in~\cite{landmann2019non}. We then turn to the characterization of the two different phases of the problem and start in Section~\ref{Sec: Unsolvable} with the unsolvable phase. Section~\ref{Sec: Solvable} contains the corresponding analysis of the solvable phase. Finally, we summarize our findings in Section~\ref{Sec: Conclusion}.

\section{Problem and notation}\label{Sec: Problem and notation}
\subsection{Large systems of random linear equations}
We study a system of $N$ random linear equations
\begin{equation}
\hat{a}^T\mathbf{x}=\mathbf{b}, \, \, \mathbf{x} \geq \mathbf{0},
\label{Equ: Linear Equation System}
\end{equation}
for $S$ unknowns $x_\mu$ where the crucial qualification $\mathbf{x} \geq \mathbf{0}$ stands for $x_\mu\geq 0$ for all $\mu=1,\dots, S$. We are hence only looking for solution vectors with {\em non-negative} components. The $S \times N$ matrix $\hat{a}$ has independent random entries $a_{\mu i}$ drawn from a Gaussian distribution with average $A$ and variance $\sigma^2$,
\begin{equation}
\langle a_{\mu i}\rangle =A,\qquad \langle (a_{\mu i}-A)^2\rangle=\sigma^2,
\end{equation} 
and $\hat{a}^T$ denotes its transpose. The components $b_i, i=1,\dots, N$ of the inhomogeneity vector $\mathbf{b}$ are drawn from a Gaussian distribution with average $B$ and variance $\gamma^2/N$,
\begin{equation}\label{Equ: Statistics of b}
\langle b_i \rangle =B, \qquad \langle (b_i-B)^2\rangle =\frac{\gamma^2}{N}.
\end{equation}  
Here and in the following the brackets $\langle \dots\rangle$ denote the average over the  $a_{\mu i}$ and $b_i$. 
We are always interested in the large system limit $N\to \infty$ with $S=\alpha N$ and $\alpha=\mathcal{O}(1)$. Note the scaling of the variance $\langle (b_i-B)^2\rangle \sim 1/N$ with the system size. It is  important for a controlled limit $N\to\infty$ in the subsequent calculations.

\subsection{Solvable and unsolvable phase}
For $N \rightarrow \infty$ the system (\ref{Equ: Linear Equation System}) exhibits a phase transition from a phase where a solution almost always exists to one where typically no such solution can be found. This line depends on the ratio $\alpha$ between the number of variables and the number of equations and on the parameters $A, \, B, \, \sigma, \, \gamma $ of the probability distributions involved. The analytic determination of the transition line was performed in \cite{landmann2019non}. To make this paper self-contained and to set the stage for the ensuing calculations the main steps of the corresponding calculations are sketched in Sec.~\ref{Sec: Determination of critical line}. For more details the reader is referred to the original paper.

The central aim of the present work is to characterize the two phases of system (\ref{Equ: Linear Equation System}). To this end we first need to find quantities able to describe relevant properties of the system away from criticality.

In the unsolvable phase no solution $\mathbf{x}$ to \eqref{Equ: Linear Equation System} exists and it is intuitive to ask ``how far'' the system is from having a solution. This somewhat vague concept can be quantified by the minimal residual norm:
\begin{equation}\label{Equ:r in the x picture}
r:=\min_{\mathbf{x},\mathbf{x}\geq \mathbf{0}} \left\|\hat{a}^T\mathbf{x}-\mathbf{b}\right\|, 
\end{equation} 
where $\|\dots\|$ denotes the usual quadratic norm of vectors. As long as no solution $\mathbf{x}$ exists $r$ is non-zero. By approaching the solvable phase $r$ should decrease and eventually tend to zero at the transition. We are interested in the detailed behavior of $r$ as a function of $\alpha,\, A, \, B, \, \sigma,$ and $\gamma$.

In the solvable phase the situation is complementary. Now a solution always exists. It is then natural to ask for the robustness or flexibility of the system: How strongly can it be changed or perturbed while still remaining  solvable? There are several possibilities to phrase this idea mathematically. One is to determine the {\em minimal}  number of rows $\mathbf{a}_\mu$ and corresponding variables $x_\mu$ by which the system (\ref{Equ: Linear Equation System}) has to be reduced such that it is rendered unsolvable. We will discuss related measures in section~\ref{Sec: Solvable}.

\begin{figure}[]
	\centering
	\includegraphics[width= 0.4\linewidth]{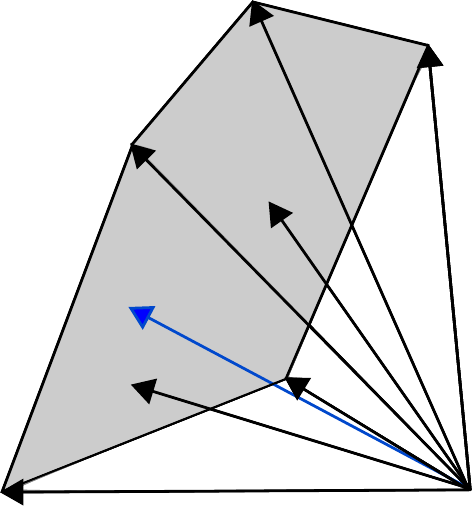}
	\caption{The geometry of problem (\ref{Equ: Linear Equation System}) in the solvable phase. The inhomogeneity vector $\mathbf{b}$ (blue) belongs to the cone spanned by non-negative linear combinations \eqref{eq:defcone} of the row vectors $\mathbf{a}_\mu$ (black). Correspondingly, system (\ref{Equ: Linear Equation System}) has a solution $\mathbf{x}$. For the situation in the unsolvable phase see Fig.~\ref{Fig: geometric interpretation r}.}
	\label{Fig: solvable phase}
\end{figure}

\section{Farkas' Lemma}\label{Sec: Farkas}

It is very useful to map the original problem (\ref{Equ: Linear Equation System}) to a dual one with the help of Farkas' lemma~\cite{farkas1902theorie}. To this end we consider what is called the {\em non-negative cone} of row vectors $\mathbf{a}_\mu$ of matrix $\hat{a}$. This cone is spanned by all linear combinations 
\begin{equation}\label{eq:defcone}
c_1 \mathbf{a}_1+c_2 \mathbf{a}_2+\dots + c_{\alpha N} \mathbf{a}_{\alpha N}
\end{equation} 
of these row vectors with non-negative coefficients $c_\mu$, cf.~Fig.~\ref{Fig: solvable phase}. Clearly, if  $\mathbf{b}$ falls into this cone, \eqref{Equ: Linear Equation System} has a solution whereas no such solution exists when $\mathbf{b}$ lies outside the cone. In the latter case, however, a hyperplane must exist that separates $\mathbf{b}$ from the cone. If we denote the normal of this hyperplane by $\mathbf{y}\in\mathbb{R}^N$ we may hence state that 
{\em either} a solution $\mathbf{x}$ to 
\begin{equation}\label{eq:nqw}
\hat{a}^T\mathbf{x}=\mathbf{b} \,  \, \text{with} \, \,\mathbf{x \geq 0},
\end{equation} 
exists {\em or} we may find a vector $\mathbf{y}$ satisfying  
\begin{equation}\label{eq:y} 
\hat{a} \mathbf{y} \geq \mathbf{0} \, \, \text{and}  \, \,  \mathbf{b}\cdot\mathbf{y} < 0.
\end{equation} 
This duality is the pivotal point of our analysis. As we will show in the next sections \eqref{eq:y} may be analyzed using the replica method.

\section{Determination of the critical line}\label{Sec: Determination of critical line}
Let us first see how we can determine the critical line of the system by studying the inequalities (\ref{eq:y}). If a vector $\mathbf{y}$ exists which fulfills all inequalities the original system (\ref{Equ: Linear Equation System}) does not have solution. If there is no such vector $\mathbf{y}$ a non-negative solution $\mathbf{x}$ exists.

To get rid of the trivial degeneracy of solutions $\mathbf{y}$ implied by $\mathbf{y} \rightarrow \omega \mathbf{y}$ for any positive $\omega$ we first require
\begin{equation}\label{eq:sphconstr}
\|\mathbf{y}\|^2=\sum_i y_i^2=N.
\end{equation} 
Next we define the fractional volume $\Omega (\hat{a},\mathbf{b})$ of all vectors $\mathbf{y}$ which fulfill (\ref{eq:y}) for given $\hat{a}$ and $\mathbf{b}$:
\begin{widetext}
\begin{align}
\Omega (\hat{a},\mathbf{b}):=
\frac{\int^\infty_{-\infty} \prod_i dy_i\, \delta \left(\sum_i y_i^2-N\right)
	\Theta \left(-\frac{1}{\sqrt{N}}\sum_i  b_i y_i \right)
	\prod_\mu\Theta \left( \frac{1}{\sqrt{N}} \sum_i a_{\mu i} y_i \right)  }{\int^\infty_{-\infty} \prod_i dy_i \,\delta(\sum_i y_i^2-N)}.
\label{Equ: Volume of solutions}
\end{align}
\end{widetext}
Here $\Theta$ denotes Heaviside-functions and their arguments have been scaled such that their typical values remain $\mathcal{O}(1)$ for $N\to \infty$. 

\begin{figure}[]
	\centering
	\includegraphics[width=0.49\linewidth]{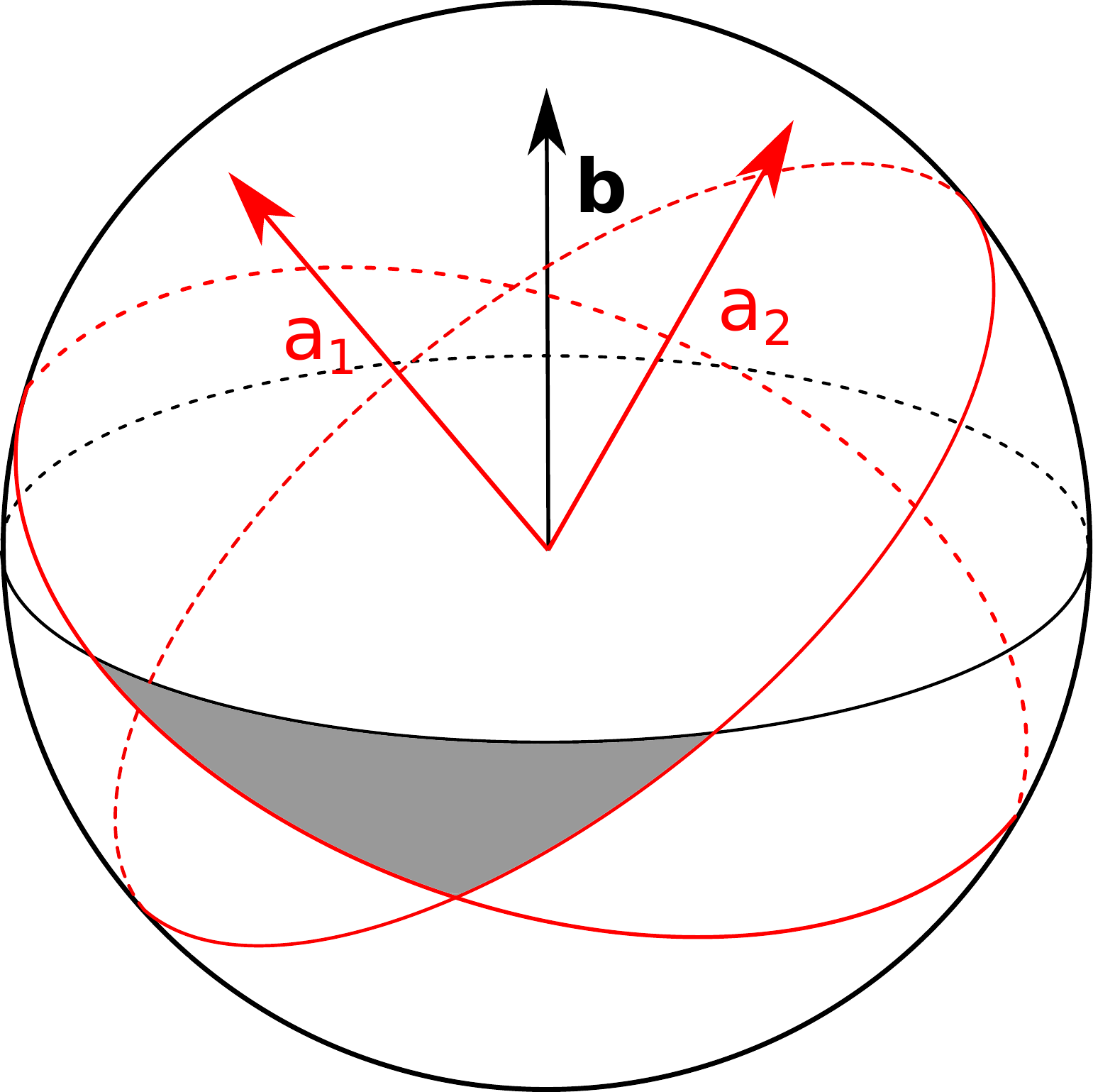}
	\caption{Volume of allowed solutions $\mathbf{y}$ of (\ref{eq:y}) (gray shaded area). All vectors $\mathbf{y}$ lie on a sphere of radius $\sqrt{N}$. The inequality $\mathbf{y} \cdot \mathbf{b}<0$ constrains the vectors to lie on the lower hemisphere with respect to $\mathbf{b}$. At the same time, they need to have a positive scalar product with all row vectors $\mathbf{a}_\mu$.}
	\label{Fig: volume of solutions y}
\end{figure}

The geometric interpretation of $\Omega$ is depicted in Figure~\ref{Fig: volume of solutions y}. Due to \eqref{eq:sphconstr} all vectors $\mathbf{y}$ lie on a sphere with radius $\sqrt{N}$. The inequality $-\mathbf{y} \cdot \mathbf{b}<0$ in \eqref{eq:y} constrains them to the lower hemisphere with respect to $\mathbf{b}$. At the same time, all $\mathbf{y}$ need to have a non-negative scalar product with every row vector of $\hat{a}$. The more variables the system (\ref{Equ: Linear Equation System}) has, the more constraints need to be fulfilled by $\mathbf{y}$. Therefore, the fractional volume of solutions (gray shaded area) shrinks, as the number of variables is increased. When the fractional volume becomes zero, there is no $\mathbf{y}$ satisfying all the inequalities. This implies that a solution to the system (\ref{Equ: Linear Equation System}) exists.

The value of $\Omega (\hat{a},\mathbf{b})$ depends on the specific choice of the matrix $\hat{a}$ and the vector $\mathbf{b}$. Rather than to pinpoint the fractional volume for each individual realization of the randomness we are interested in its {\em typical} value under the distributions defined in section~\ref{Sec: Problem and notation}. 
Since $\Omega$ is dominated by a {\em product} of many independent random terms, we cannot expect its average to give a good estimate for its typical value. Instead, this typical value is given by
\begin{equation}
\Omega_\mathrm{typ}\simeq \exp\left(\langle \log \Omega\rangle\right).
\end{equation} 
For given statistical parameters $A, \, B, \, \sigma, \, \gamma $ the critical line of the transition is determined by finding the value of $\alpha$ for which the {\em typical value} of the volume $\Omega$ becomes zero. Therefore, our quantity of interest is the entropy:
\begin{equation}\label{defS}
S(\alpha,A,B,\sigma,\gamma):=\lim_{N\to\infty}\frac{1}{N}
\langle\log \Omega(\hat{a},\mathbf{b})\rangle.
\end{equation}
For its determination we use the replica trick \cite{mezard1987spin} which is based on the identity: 
\begin{equation}\label{replica}
\left< \log  \Omega \right>=\lim_{n\rightarrow 0}\frac{ \left<\Omega^n\right>-1}{n}.
\end{equation}
The calculation is feasible for $n\in \mathbb{N}$. The result has then to be continued to the real $n$ in order to accomplish the crucial limit $n\to 0$. For $n\in \mathbb{N}$ we have
\begin{align}
&\Omega^n  (\hat{a},\mathbf{b})= \nonumber
\int^\infty_{-\infty} \prod_{i,a} \frac{dy^a_i}{\sqrt{2 \pi e}} \prod_a \delta \left(\sum_i (y^a_i)^2-N \right)
\\
&
\prod_{\mu,a} \Theta \left( \frac{1}{\sqrt{N}} \sum_i a_{\mu i} y^a_i \right)\prod_a \Theta \left(-\frac{1}{\sqrt{N}}\sum_i b_i y^a_i \right).
\label{Equ:replica Volume of solutions}
\end{align}
Here, $a$ is the replica index, which runs from $1 \dots n$, and the denominators $\sqrt{2\pi e}$ account for the denominator in \eqref{Equ: Volume of solutions}. Replacing the $\delta$- and $\Theta$-functions by their integral representations the arising integrals can be decoupled by introducing the order parameters:
\begin{equation}\label{eq:defop}
m^a=\frac{1}{\sqrt{N}}\sum_i y^a_i\quad\mathrm{and}\quad 
q^{ab}=\frac{1}{N}\sum_i y^a_i y^b_i\quad\mathrm{for}\; a<b.
\end{equation} 

In the limit $N\rightarrow \infty$ the integrals may then be calculated by the saddle-point method. Since the solution space of (\ref{eq:y}) is connected the replica-symmetric ansatz for the saddle-point is expected to yield correct results:
\begin{align}
m^a= \, m, \quad q^{ab}= \, q, \quad\mathrm{for}\; a< b.
\end{align}
Some of the resulting saddle-point equations are algebraic and can be used to eliminate the corresponding variables. With the abbreviations 
\begin{equation}\label{defDH}
 Dt:=\frac{dt}{\sqrt{2 \pi}}e^{-t^2/2},\qquad H(x):=\int_x^\infty \!\! Dt
\end{equation} 
as well as 
\begin{equation}\label{deflambda}
 \kappa:=\frac{m A}{\sigma},\qquad \lambda:= \frac{\sigma B}{\gamma A}
\end{equation} 
we finally arrive at \cite{landmann2019non}
\begin{align}\label{Sres}
S(\alpha,\lambda)&=\mathrm{extr}_{q,\kappa}\Big[\frac{1}{2}\log(1-q)+\frac{q}{2(1-q)} \nonumber
\\
&-
\frac{\lambda^2}{2}\frac{\kappa^2}{1-q}
+\alpha\int \!\! Dt\, \log H \Big(\frac{\sqrt{q}\,t-\kappa}{\sqrt{1-q}}\Big)\Big].
\end{align} 
The order parameter $q$ characterizes the typical overlap between two different solutions $\mathbf{y}$ from the solution space. It varies from $q=0$ for $\alpha=0$ to $q\to 1$ close to the phase transition. To determine the transition line we may hence reduce Eq.~\eqref{Sres} to its most divergent terms for $q\to 1$. In this limit we have 
\begin{align}
\int \!\! Dt\, \log H \Big(\frac{\sqrt{q}\,t-\kappa}{\sqrt{1-q}}\Big) \nonumber
&\sim -\frac{1}{2(1-q)}\int_\kappa^\infty \!\! Dt\,(t-\kappa)^2
\\\label{defI}
&=:-\frac{1}{2(1-q)}\, I(\kappa).
\end{align} 
and correspondingly get 
\begin{align}
S(\alpha_c,\lambda)\sim &\, \mathrm{extr}_{q,\kappa} \, \Big[\frac{1}{2(1-q)} \nonumber
\\
-&\frac{\lambda^2}{2}\frac{\kappa^2}{(1-q)}-\frac{\alpha_c}{2(1-q)} I(\kappa)\Big].
\end{align} 

The critical value $\alpha_c$ of $\alpha$ is determined by the remaining saddle-point equations for $q$ and $\kappa$:
\begin{align}
1-\lambda^2 \kappa^2&=\alpha_c I(\kappa), \label{h4}
\\
\alpha_c H(\kappa)&=1. \label{h5}
\end{align}
Note that in this result the statistical properties of $a_{\mu i}$ and $b_i$ as specified by the parameters $A, \, B, \, \sigma, \gamma$ only arise in form of $\lambda$ defined in \eqref{deflambda} which gives the scaled ratio of the relative variances of $a_{\mu i}$ and $b_i$.

\begin{figure}[]
	\centering
	\includegraphics[width=0.9 \linewidth]{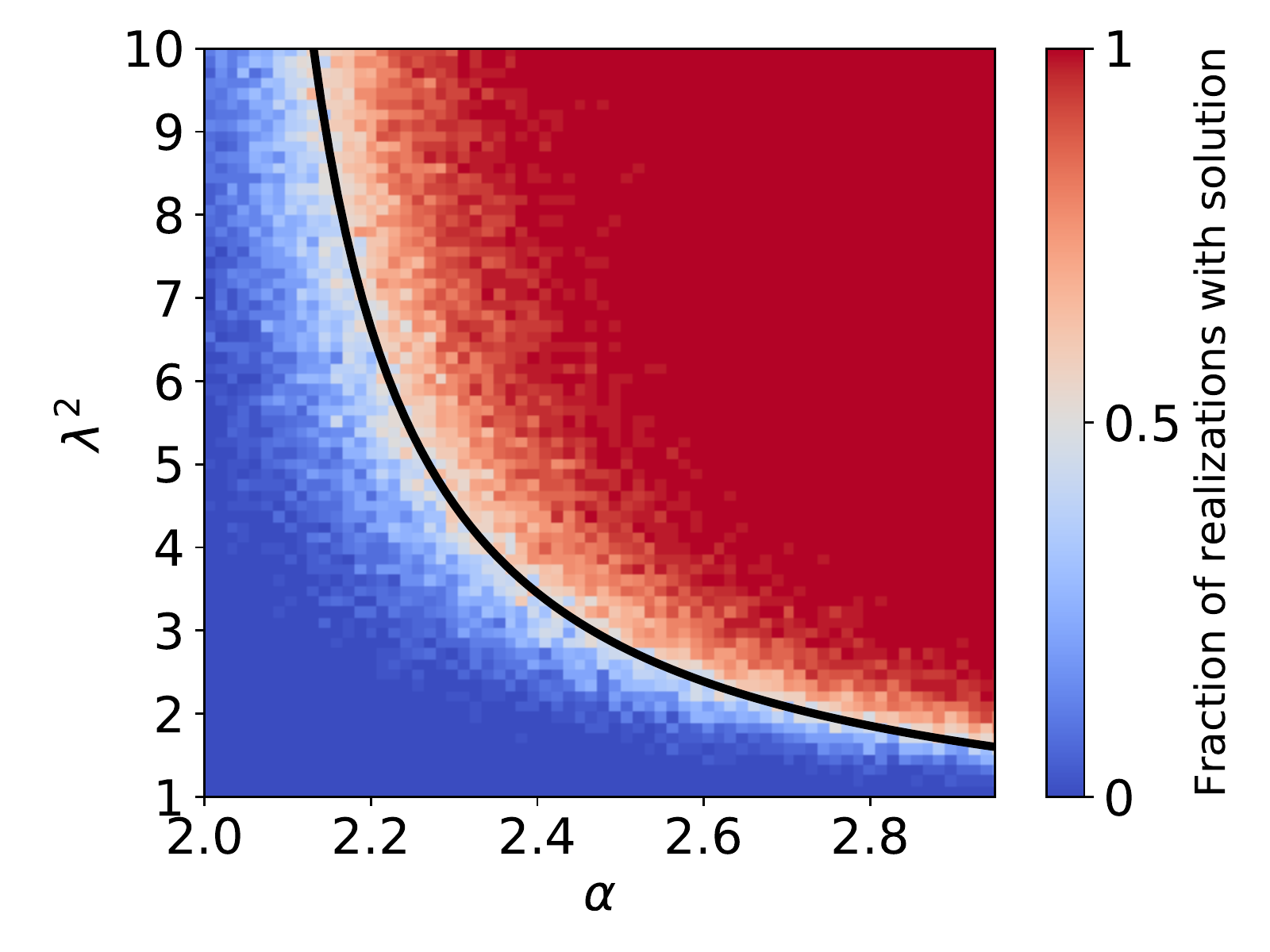}
	\caption{The transition from the unsolvable to the solvable phase in dependence of the statistical properties of the random variables described by $\lambda$ and the ratio  $\alpha$ between the number of variables and the number of equations. The color indicates the fraction of randomly drawn systems (\ref{Equ: Linear Equation System}) for which a non-negative solution could be found numerically. The black line shows the analytical prediction of the critical line given by Eqs.~(\ref{h4}),(\ref{h5}). Parameters: $N=300$, every data point shows the average over 50 realizations.}
	\label{Fig: phase transition}
\end{figure}

Figure~\ref{Fig: phase transition} shows the critical line (black) given by (\ref{h4}), (\ref{h5}) in comparison with simulation results. The color indicates the fraction of randomly drawn systems for which a non-negative solution could be found using a non-negative least squares solver  \setcounter{footnote}{0}\footnote{We used the non-negative least squares solver nnls from the scipy.optimize package in Python.}. There is good agreement between the analytical prediction and the numerics. 

\section{The unsolvable phase}\label{Sec: Unsolvable}
\begin{figure}[]
	\centering
	\includegraphics[width=0.6 \linewidth]{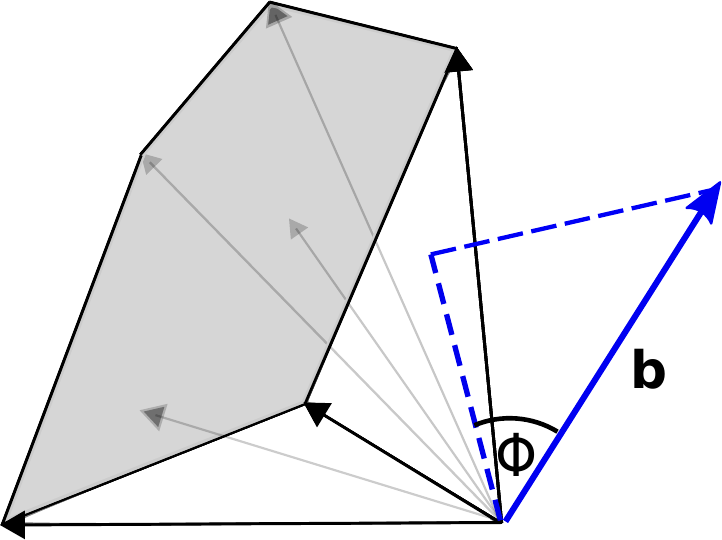}
	\caption{In the unsolvable phase the inhomogeneity vector $\mathbf{b}$ lies outside the cone of possible non-negative linear combinations $\hat{a}^T  \mathbf{x} = \sum_{\mu}^{S} x_\mu \mathbf{a}_\mu$. The minimal residual norm $r=\min_{\mathbf{x},\mathbf{x}\geq \mathbf{0}} \left\|\hat{a}^T\mathbf{x}-\mathbf{b}\right\|$, is realized by a linear combination which lies on the face of the cone closest to $\mathbf{b}$. Therefore, it is completely determined by the smallest angle $\phi_{\min}$ between the cone and $\mathbf{b}$. }
	\label{Fig: geometric interpretation r}
\end{figure}

In the unsolvable phase of problem \eqref{Equ: Linear Equation System} no solution vector $\mathbf{x}$ may be found. In section~\ref{Sec: Problem and notation} we proposed the residual norm $r$ defined in~\eqref{Equ:r in the x picture} as a measure of how far the system is from having a solution. We will show now how to determine the typical value of $r$ from the dual problem \eqref{eq:y}. To this end it is useful to first give $r$ a geometrical interpretation, see Fig.~\ref{Fig: geometric interpretation r}. In the unsolvable phase the inhomogeneity vector $\mathbf{b}$ lies outside the cone of possible non-negative linear combinations $\hat{a}^T \mathbf{x}$. The best approximation to $\mathbf{b}$ by a vector $\hat{a}^T \mathbf{x}$ of the cone lies on a face of this cone and realizes the smallest possible angle $\phi_{\min}$ between $\mathbf{b}$ and the surface of the cone. We hence get  $r=b \sin(\phi_{\min})$ with $b=\|\mathbf{b}\|$.

It remains to find a way to determine $\phi_{\min}$ from the dual problem involving the vectors $\mathbf{y}$. Since there is no solution to equation (\ref{Equ: Linear Equation System}) there is at least one and in fact several hyperplanes separating the cone from $\mathbf{b}$. The angle between these hyperplanes and $\mathbf{b}$ ranges from 0 (when the hyperplane contains $\mathbf{b}$) to $\phi_{\min}$ (when the hyperplane contains the face of the cone closest to $\mathbf{b}$). Any hyperplane having an angle with $\mathbf{b}$ larger than $\phi_{\min}$ violates at least one of the constraints in $\hat{a}  \mathbf{y} \geq \mathbf{0}$. Thus, we can determine $\phi_{\min}$ by finding the hyperplane fulfilling the constraints given by Farkas' lemma and making the {\em largest possible} angle with $\mathbf{b}$. In terms of normal vectors $\mathbf{y}$ this is equivalent to 
\begin{equation}\label{Equ: r=max_y}
r=\max_\mathbf{y} \,\left( -\frac{1}{\sqrt{N}} \mathbf{y} \cdot \mathbf{b}\right)
\end{equation} 
where the maximum is over all vectors $\mathbf{y}$ fulfilling
\begin{equation}
 \hat{a}\mathbf{y} \geq \mathbf{0} \,\,\, , \,\,\, \mathbf{b} \cdot \mathbf{y} < 0, \,\,\, \|\mathbf{y}\|^2=N.
\end{equation} 

\begin{figure}[]
\centering
\includegraphics[width=0.49 \linewidth]{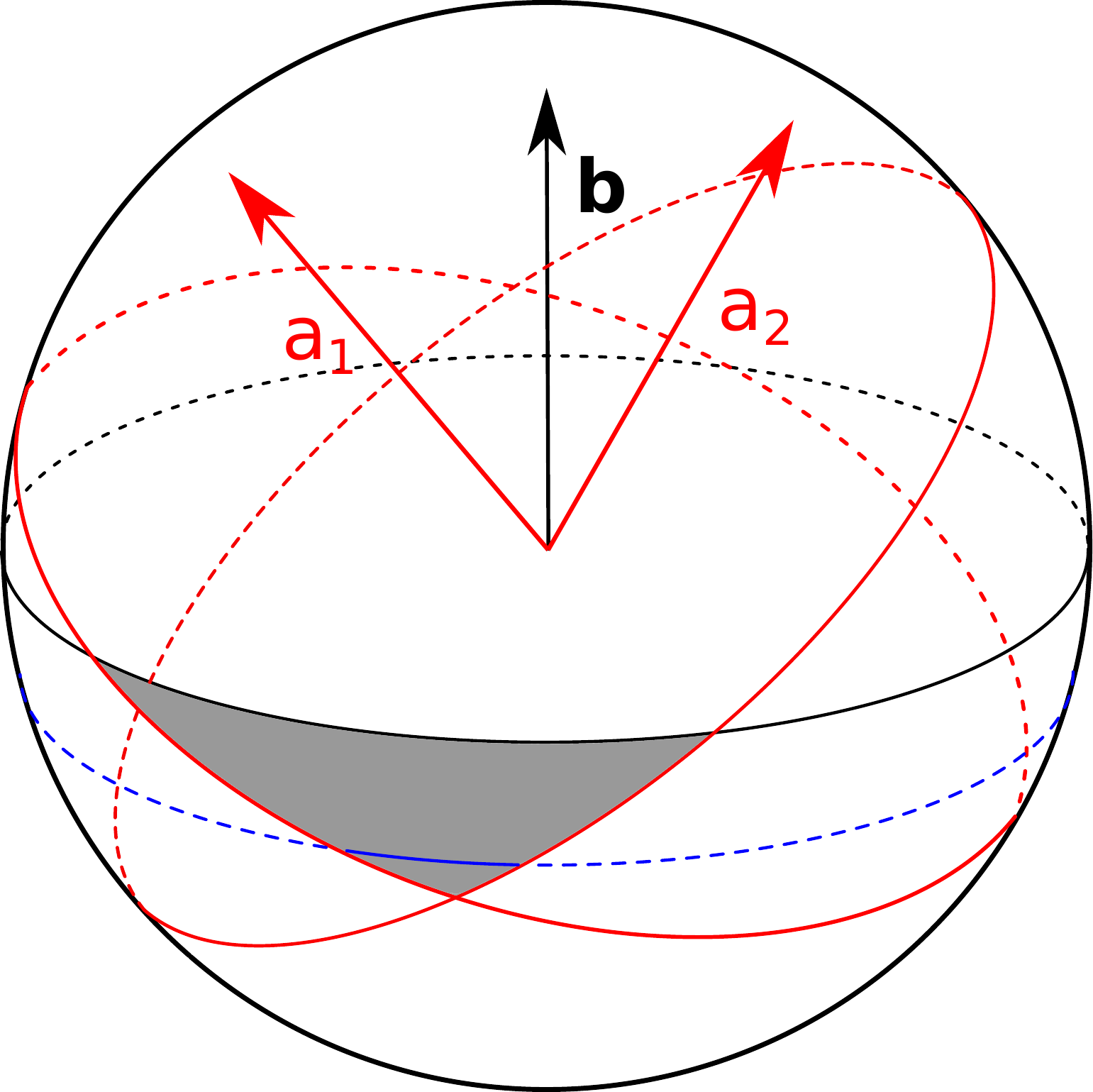}
\caption{Solution volume $\Omega$ (gray shaded) as in Fig.~\ref{Fig: volume of solutions y} together with the locus of vectors $\mathbf{y}$ making a definite angle with $\mathbf{b}$ (blue line), cf. Eq.~\eqref{defeta}. At the maximum value of this angle the line hits the ``southern tip'' of the gray area.}
\label{Fig: procedure to determine r}
\end{figure}

This formulation allows to determine the typical value of $r$ by a slight extension of the methods described in section~\ref{Sec: Determination of critical line}, see also \cite{engel1993systems} for a related situation. The procedure is best explained by comparing Figs.~\ref{Fig: volume of solutions y} and \ref{Fig: procedure to determine r}. In order to determine $r$ we have to find the maximal value $\eta_\mathrm{max}$ of 
\begin{equation}\label{defeta}
 \eta:=-\frac{1}{\sqrt{N}} \mathbf{y} \cdot \mathbf{b}
\end{equation} 
in the shaded solution space $\Omega$. We proceed as follows: We first fix a value of $\eta $ thereby constraining the vectors $\mathbf{y}$ to lie on a given latitude shown by the dotted blue line in Fig.~\ref{Fig: procedure to determine r}. The part of this line inside $\Omega$ (full blue line) represents those vectors $\mathbf{y}$ that belong to $\Omega$ and make the required angle with $\mathbf{b}$. As $\eta$ increases the full blue line gets shorter and shorter. It shrinks to a point at the ``most southern tip'' of the solution space when the maximal possible value $\eta_\mathrm{max}$ of $\eta$ has been reached. We may therefore determine $\eta_\mathrm{max}$ by calculating the modified fractional volume $\tilde{\Omega}(\hat{a},\mathbf{b},\eta)$ including the constraint \eqref{defeta} and looking for the value of $\eta$ for which it goes to zero.

The modified fractional volume $\tilde{\Omega}(\hat{a},\mathbf{b},\eta)$ is obtained from $\Omega(\hat{a},\mathbf{b})$ as defined in (\ref{Equ: Volume of solutions}) by the mere replacement
\begin{equation}
 \Theta \left(-\frac{1}{\sqrt{N}}\sum_i  b_i y_i\right)\;\longrightarrow\;
 \delta \left(-\frac{1}{\sqrt{N}}\sum_i  b_i y_i - \eta\right).
\end{equation} 
The explicit calculations are hence rather similar to Section~\ref{Sec: Determination of critical line}. They yield the following expression for the corresponding entropy
\begin{align}\label{Equ: Final Entropy determination of r}
&\tilde{S}(\alpha,\lambda,\eta/\gamma)=\lim_{N\to\infty}\frac{1}{N}
\langle\log \tilde{\Omega}(\hat{a},\mathbf{b},\eta)\rangle= \nonumber 
\\
&\text{extr}_{q,\kappa}\Big\{\frac{1}{2}\log(1-q)+\frac{q}{2(1-q)}-\frac{\left(\eta/\gamma+\kappa \lambda\right)^2}{2(1-q)} \nonumber
\\
&+\alpha\int \!\! Dt\, \log H \Big(\frac{\sqrt{q}\,t-\kappa}{\sqrt{1-q}}\Big) +C\Big\},
\end{align}
where $C$ is an irrelevant constant arising from the different normalizations of $\Omega(\hat{a},\mathbf{b})$ and $\tilde{\Omega}(\hat{a},\mathbf{b},\eta)$.

The limit $\tilde{\Omega} \rightarrow 0$ is again accompanied by $q \to 1$. Keeping only the most divergent terms we find
\begin{align}
\tilde{S}(\alpha,\lambda,\,&\eta_{\max}/\gamma)=\text{extr}_{q,\kappa}\Big\{\frac{1}{2(1-q)} \nonumber
\\
&-\frac{\left(\eta_{\max}/\gamma+\kappa \lambda\right)^2}{2(1-q)} 
-\frac{\alpha}{2(1-q)} \, I\left(\kappa\right) \Big\},
\end{align}
with the corresponding saddle-point equations
\begin{align}\label{Equ: Saddle point equations for r}
1-(\eta_{\max}/\gamma+\kappa \lambda)^2=& \, \alpha \,  I\left(\kappa\right), \nonumber
\\
\kappa \lambda (\eta_{\max}/\gamma+\kappa  \lambda)=& \, \alpha \left(H\left(\kappa\right)-I\left(\kappa\right)\right).
\end{align}
Here $H$ and $I$ denote the functions defined in \eqref{defDH} and \eqref{defI}, respectively. From the numerical solutions to \eqref{Equ: Saddle point equations for r} we get our final result $r=\eta_{\max}$. 

\begin{figure}[]
	\centering
	\includegraphics[width= 0.9\linewidth]{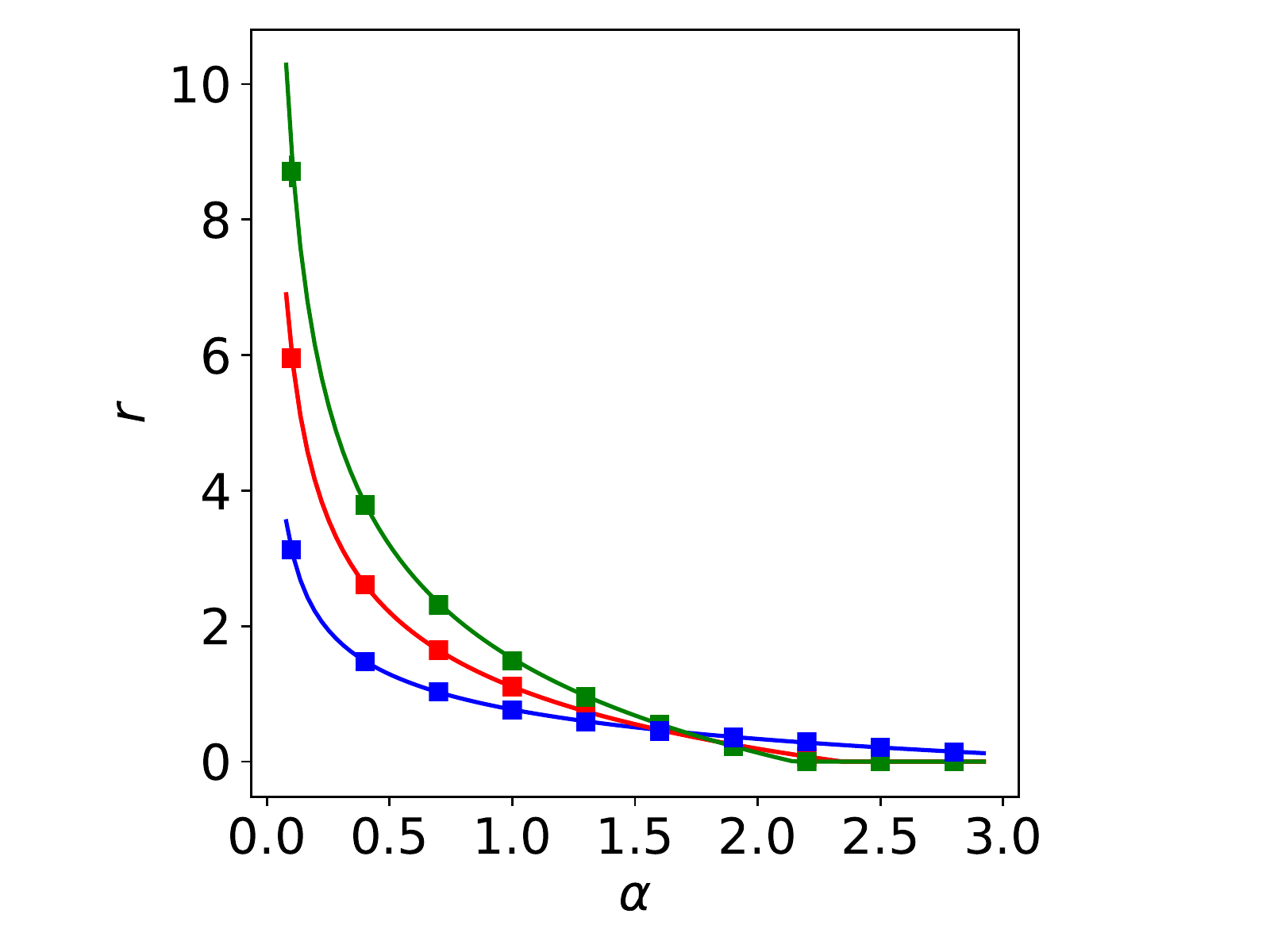}
	\caption{Comparison of analytical results and numerical simulations for the residual error $r$ in the unsolvable phase. The standard deviations of the numerical results are smaller than the symbol size. Parameters: $A=B=\gamma=1$, $\sigma=1,2,3$ (blue, red, green), $N=1000$, averaged over 10 realizations for each data point.}
	\label{Fig: results for r}
\end{figure}

Figure~(\ref{Fig: results for r}) compares results for $r$ obtained in this way with numerical simulations for systems of size $N=1000$. The data points were generated by averaging over 10 realizations of the randomness. For each realization the minimal residual norm $r$ of the random system Eq.~(\ref{Equ: Linear Equation System}) was  determined using a least squares solver \cite{Note1}. There is very good agreement between numerics and analytical results. The qualitative behavior is as expected: In the unsolvable phase, $\alpha < \alpha_c$, the minimal residual norm is nonzero. It monotonically decreases as the system approaches $\alpha_c$ to eventually reach zero at $\alpha=\alpha_c$.  

\section{The solvable phase}\label{Sec: Solvable}

Similar to the previous section we now try to characterize the solvable phase of (\ref{Equ: Linear Equation System}) by using the dual picture~(\ref{eq:y}). In the solvable phase, it is not possible to find a vector $\mathbf{y}$ which fulfills all inequalities in (\ref{eq:y}) simultaneously. We may then ask for the best $\mathbf{y}$ that violates the inequalities $\hat{a}  \mathbf{y} \geq \mathbf{0}$ the least. To formalize the idea let us define a cost function
\begin{equation}\label{defDelta}
v(\mathbf{y})=\sum_{\mu} E\left(\Delta_\mu\right), \quad \Delta_\mu:=\frac{1}{\sqrt{N}}\sum_ia_{\mu i}y_i 
\end{equation}
which assigns an 'energy' $E$ to each violated constraint.

\begin{figure}[]
	\centering
	\includegraphics[width= 0.9\linewidth]{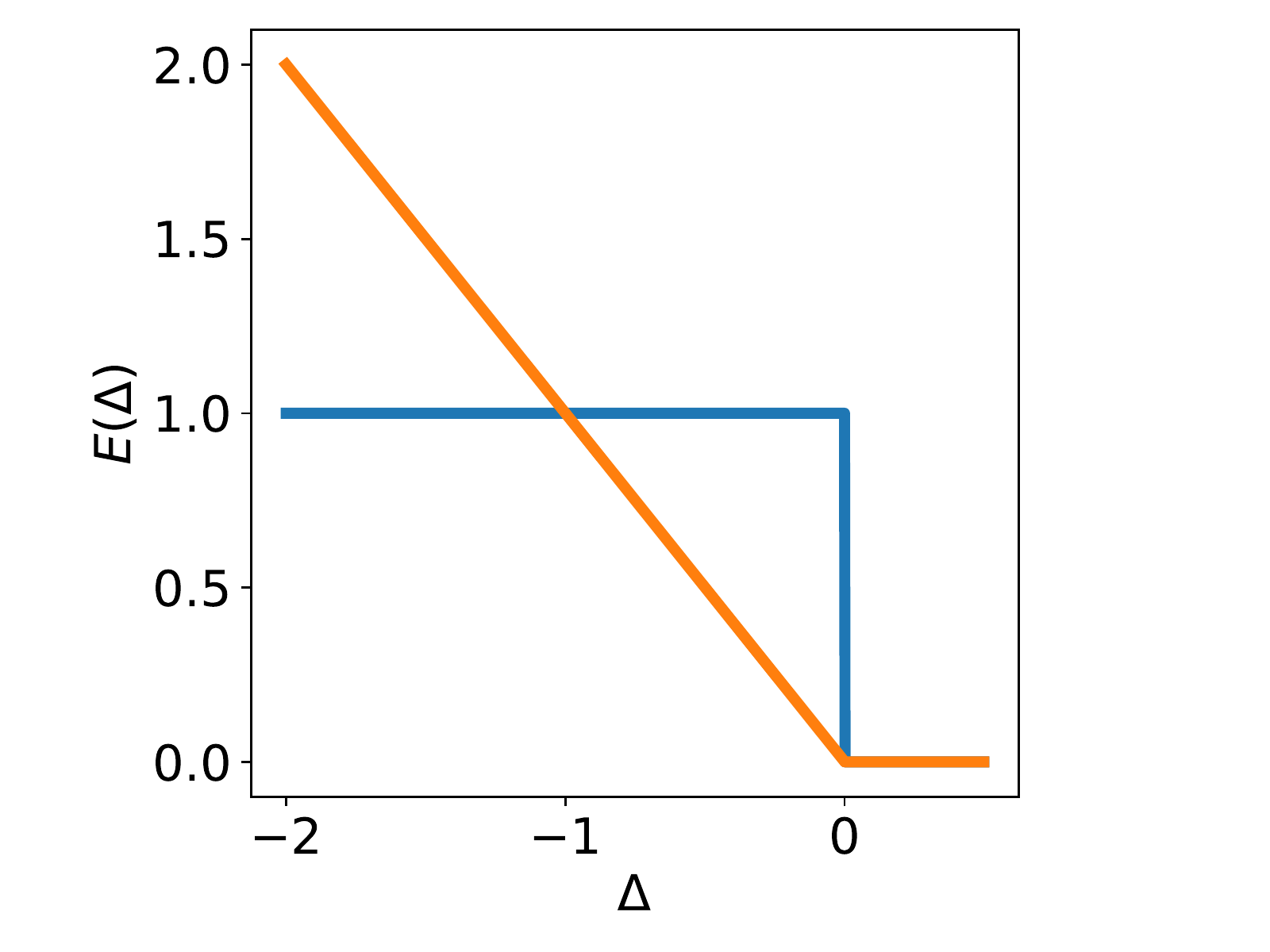}
	\caption{The two cost functions studied in this section: The step function defined in \eqref{defstep} (blue) and the ramp function \eqref{deframp} (orange).}
	\label{Fig: step vs ramp cost function}
\end{figure}

Two plausible choices for such a cost function are compared in Fig.~\ref{Fig: step vs ramp cost function}. The first one derives from the step function 
\begin{equation}\label{defstep}
 E\left(\Delta_\mu\right)=\Theta \left(-\Delta_\mu\right)
\end{equation} 
and simply counts the number of violated constraints. Its minimum over all choices of $\mathbf{y}$ hence corresponds to the minimal number of row vectors $\mathbf{a}_\mu$ that have to be eliminated from $\hat{a}$ such that a solution $\mathbf{y}$ exists. Going back to the original problem it therefore gives the minimal number of row vectors $\mathbf{a}_\mu$ and corresponding variables $x_\mu$ by which the system (\ref{Equ: Linear Equation System}) has to be reduced such that it does not possess a non-negative solution anymore. In view of Fig.~\ref{Fig: solvable phase} this number gives an estimate for the number of ``crucial'' row vectors $\mathbf{a}_\mu$ that are indispensable for the cone to contain $\mathbf{b}$. 

The second cost function builds on the ramp function
\begin{equation}\label{deframp}
  E\left(\Delta_\mu\right)=-\Theta \left(-\Delta_\mu\right)\,\Delta_\mu
\end{equation} 
and is sensitive not only to the violation of a constraint per se but also to the value of $\Delta_\mu$, i.e. to the strength of such a violation \footnote{Both functions have an equivalent in the theory of neural networks:  The step function corresponds to the Gardner-Derrida cost function \cite{GaDe}, the ramp function to the perceptron cost function \cite{griniasty1991learning}.}. Referring again to Fig.~\ref{Fig: solvable phase} its minimal value characterizes the sensitivity of the system (\ref{Equ: Linear Equation System}) to small variations in $\hat{a}$ and $\mathbf{b}$ because it indicates whether $\mathbf{b}$ lies ``well in the middle'' of the non-negative cone of the $\mathbf{a}_\mu$ or rather near to its boundary. 

It is again possible to calculate the typical minimal values of both cost functions within the replica approach outlined above. To this end we introduce an auxiliary inverse temperature $\beta$ and define the partition function 
\begin{align}\label{Eq: Partition function}
&Z(\hat{a},\mathbf{b},\beta)= \nonumber
\\
&\frac{\int \prod_i dy_i \delta\left(\sum_i y_i^2 -N\right)\Theta\left(-\frac{1}{\sqrt{N}}\sum_i b_i y_i\right) \exp\left\{-\beta v(\mathbf{y})\right\}}{\int \prod_i dy_i \delta\left(\sum_i y_i^2 -N\right)}.
\end{align}

It is instructive to compare this expression with the fractional volume, Eq.~(\ref{Equ: Volume of solutions}), which was used to examine the phase transition and the minimal residual norm in the unsolvable phase. There, only those $\mathbf{y}$ contributed to the integral which do not violate any constraint at all in $\hat{a}  \mathbf{y} \geq \mathbf{0}$, i.e., only those with $v(\mathbf{y})=0$. The entropy~\eqref{defS} was hence similar to a microcanonical ground state entropy. Complementary, Eq.~(\ref{Eq: Partition function}) is like a canonical partition function to which all $\mathbf{y}$ contribute, also those violating the constraints, albeit suppressed by the Boltzmann factor $\exp\{-\beta v(\mathbf{y})\}$. The role of the entropy in the microcanonical approach is now taken by the free energy
\begin{equation}
 F(\alpha,\lambda,\beta):=-\lim_{N\to\infty}\frac{1}{\alpha N \beta} \langle \log Z(\hat{a},\mathbf{b},\beta)\rangle.
\end{equation} 
The minimal possible average cost per equation $\langle E_\mathrm{min}\rangle$ corresponds to the ground state energy and is given as the low temperature limit of the free energy:

\begin{align}
\label{Equ: Minimal cost function derivative}
\left<E_{\min} \right>=\frac{\left<v_{\min} \right>}{\alpha N}
      = \lim_{\beta \to \infty} F(\alpha,\lambda,\beta).
\end{align}

\begin{figure*}[]
	\centering
	\includegraphics[width=0.41 \linewidth]{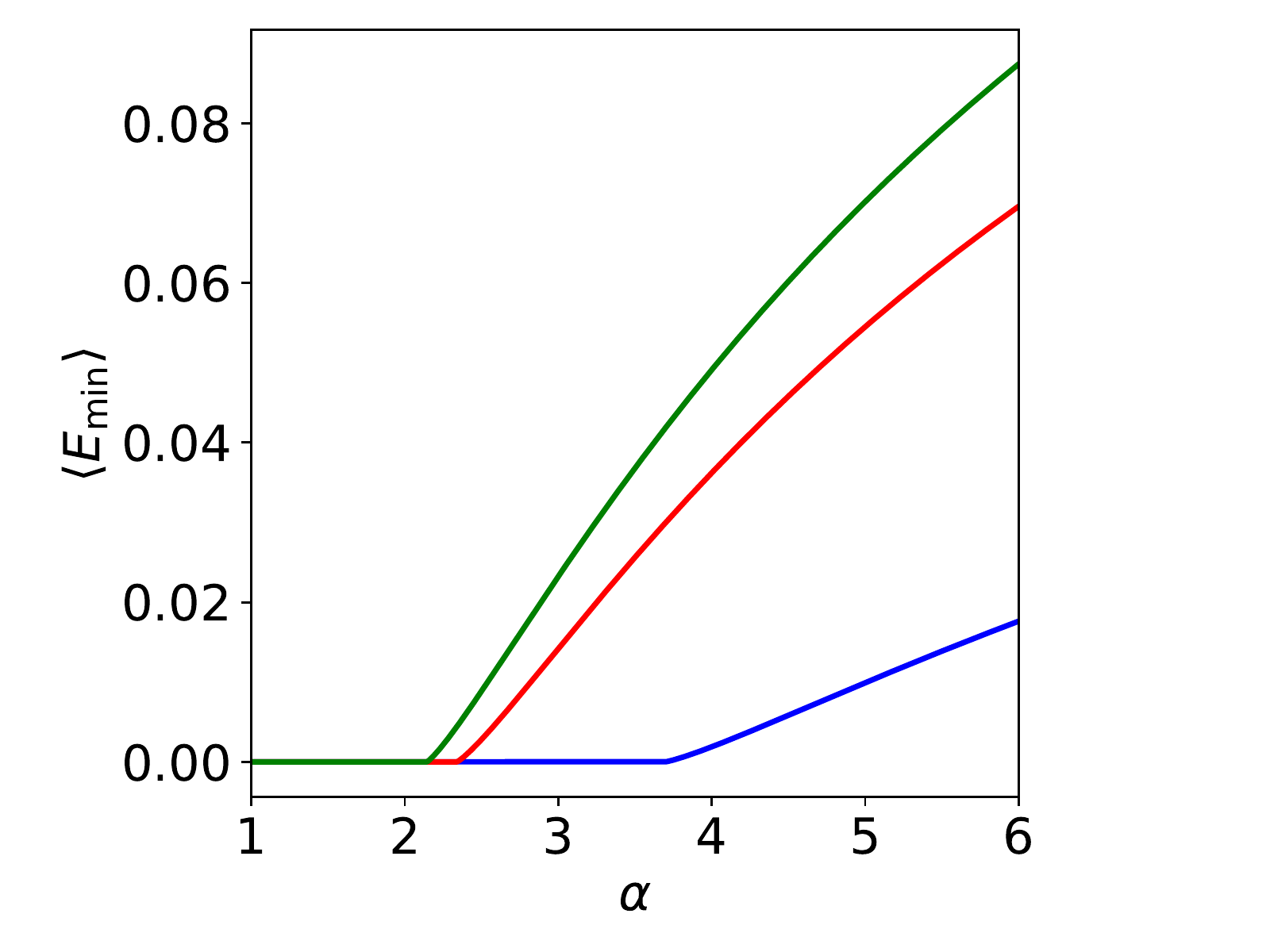}
	\hspace{0.05\linewidth}
	\includegraphics[width=0.41 \linewidth]{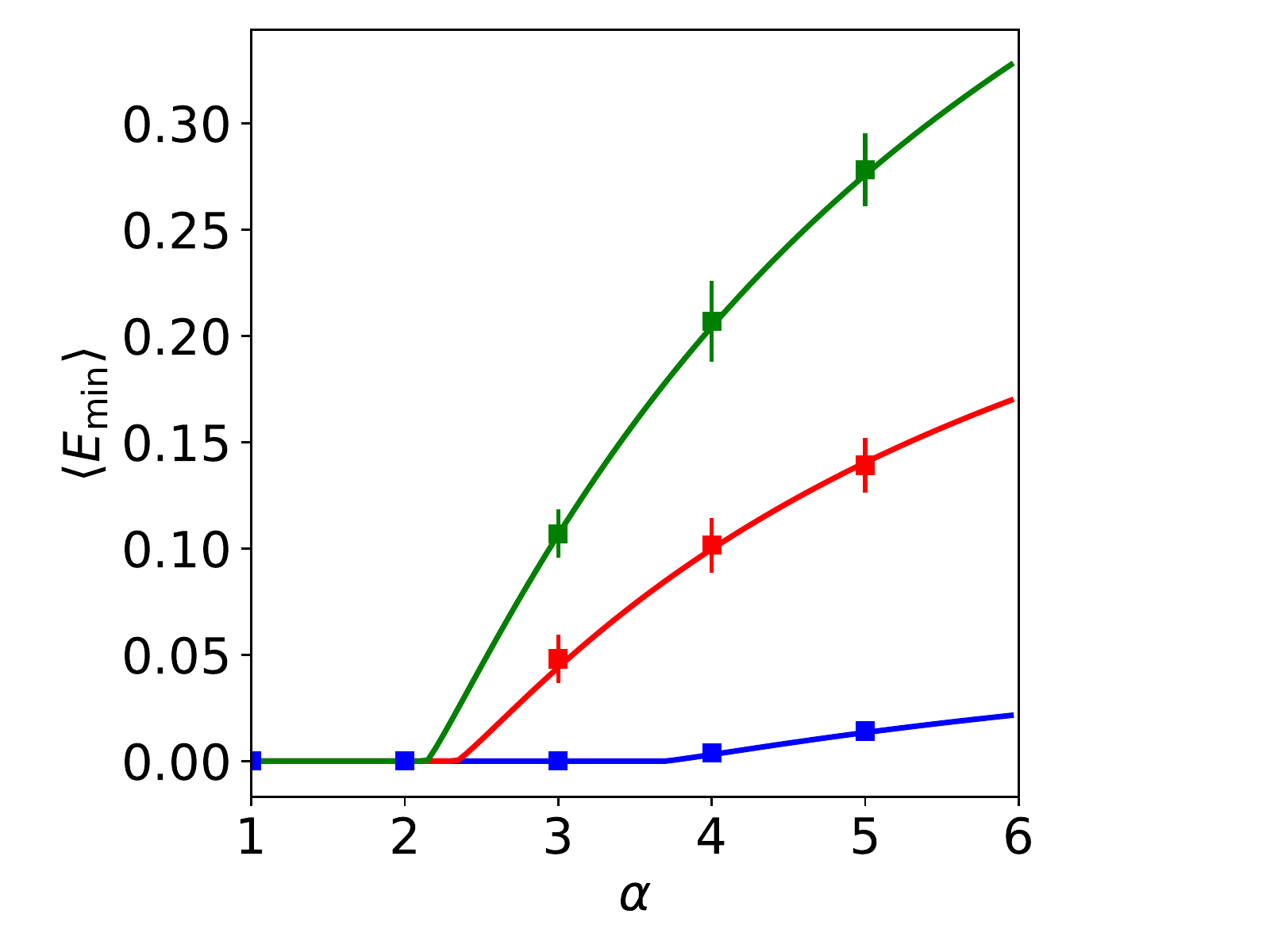}
	\caption{The lines show the typical minimal energy $\left< E_{\min} \right>$ for the step cost function (left) and for the ramp cost function (right) as given by Eq.~(\ref{Equ: Vmin Derrida}) and (\ref{Equ: Vmin Perceptron}), respectively. Parameters: $A=B=\gamma=1$, and $\sigma=1,2,3$ (blue, red, green). Each symbol of the numerical results for the ramp function was generated from systems with $N=300$   and averaged over 25 realizations. The error bars show the standard deviation. }
	\label{Fig: Minimal energies}
\end{figure*}

Its calculation is yet another variation of the one sketched in section~\ref{Sec: Determination of critical line} and proceeds along the lines of \cite{GaDe,griniasty1991learning}. The same order parameters are defined as in  (\ref{eq:defop}) and under the assumption of replica symmetry, we find 

\begin{align}
F(\alpha,\lambda,\beta) =& \nonumber
 \,\text{extr}_{q,\kappa} \Big\{-\frac{\log(1-q)}{2\alpha \beta}-\frac{q}{2\alpha \beta(1-q)}\\&+
 \frac{\lambda^2\kappa^2}{2\alpha\beta(1-q)}-\frac{1}{\beta} G_E\Big\},
\end{align}
where
\begin{align}
G_E=& \nonumber\int \!\! Dt \log\int d\Delta
\\
& \exp\left\{-\beta E\left(\Delta\right)-\frac{(\Delta/\sigma-\kappa+t \sqrt{q})^2}{2(1-q)}\right\}.
\end{align}

The limit $\beta \rightarrow \infty$ to project out the ground state is accompanied by $q \rightarrow 1$ such that \mbox{$x:=\beta (1-q)$} remains of $\mathcal{O}(1)$. In this limit, therefore, the role of the saddle-point variable $q$ is taken over by $x$. In this way $G_E$ becomes 

\begin{align}
G_E=\beta \int \!\! Dt  \left(-E\left(\Delta_0\right)-\frac{(\Delta_0/\sigma-\kappa+t )^2}{2x}\right),
\end{align}
where 
\begin{equation}\label{Equ: Minimal Delta}
\Delta_0(t)= \arg\min_\Delta \left[E(\Delta)+\frac{(\Delta_0/\sigma-\kappa+t )^2}{2x}\right].
\end{equation}
For the step cost function (\ref{defstep}) the minimum is realized by 
\begin{equation}\label{Equ: Minimal Delta step}
\Delta_0(t)=
\begin{cases}
\sigma(\kappa-t) & \text{if } t<\kappa \\
0 & \text{if } \kappa \leq t \leq \kappa+\sqrt{2x} \\
\sigma(\kappa-t) & \text{if } t > \kappa+\sqrt{2x}.
\end{cases}       
\end{equation}
The typical minimal cost per equation is then given by
\begin{align}
\langle E_\mathrm{min}\rangle 
= \text{extr}_{x,\kappa} \Big\{&-\frac{1}{2\alpha x}+\frac{\lambda^2\kappa^2}{2\alpha x}+ H\left(\kappa+\sqrt{2x}\right) 
\nonumber \\\label{eq:freeen}
&+\frac{1}{2x}\int_{\kappa}^{\kappa+\sqrt{2x}} \!\!\!\!Dt\, (t-\kappa)^2\Big\}.
\end{align}
This expression is similar to the one for the minimal fraction of misclassified patterns of a perceptron \cite{GaDe,griniasty1991learning}. Here, however, we have the additional term $\frac{\lambda^2 \kappa^2}{2x}$ and the additional extremization in $\kappa$.
\\
The saddle point equations corresponding to \eqref{eq:freeen} read
\begin{align}
\alpha \int_{\kappa}^{\kappa+\sqrt{2x}}\!\! Dt\,(t-\kappa)^2=&1-\lambda^2\kappa^2 \label{Equ: Saddle Point Equation Derrida 1}
\\
\alpha \int_{\kappa}^{\kappa+\sqrt{2x}}\!\! Dt\,(t-\kappa) = &\lambda^2 \kappa \label{Equ: Saddle point equation Derrida 2}
\end{align}
and \eqref{eq:freeen} simplifies to 
\begin{equation}\label{Equ: Vmin Derrida}
\left<E_{\min} \right>=H(\kappa+\sqrt{2x}).
\end{equation}
Note that $\kappa$ and $x$ in this expression have to be determined from the numerical solution of the saddle-point equations~\eqref{Equ: Saddle Point Equation Derrida 1} and \eqref{Equ: Saddle point equation Derrida 2}. 

Analytical results for $\left<E_{\min}\right>$ in case of the step cost function as function of $\alpha$ are shown in the left part of Figure~\ref{Fig: Minimal energies}. The overall behavior is as expected: For $\alpha$ below the critical value $\alpha_c$, the system is in the unsolvable phase, all constraints in (\ref{eq:y}) can be fulfilled and $\left<E_{\min} \right> $ is zero. For $\alpha >\alpha_c$ some of the constraints can no longer be fulfilled and $ \left< E_{\min} \right>$ starts to monotonically grow. Since the determination of $\left<E_{\min}\right>$ by simulations is computationally very demanding and not indispensable for this work we have refrained from doing so. 


\begin{figure*}
	\centering 	
	\includegraphics[width=0.41\linewidth]{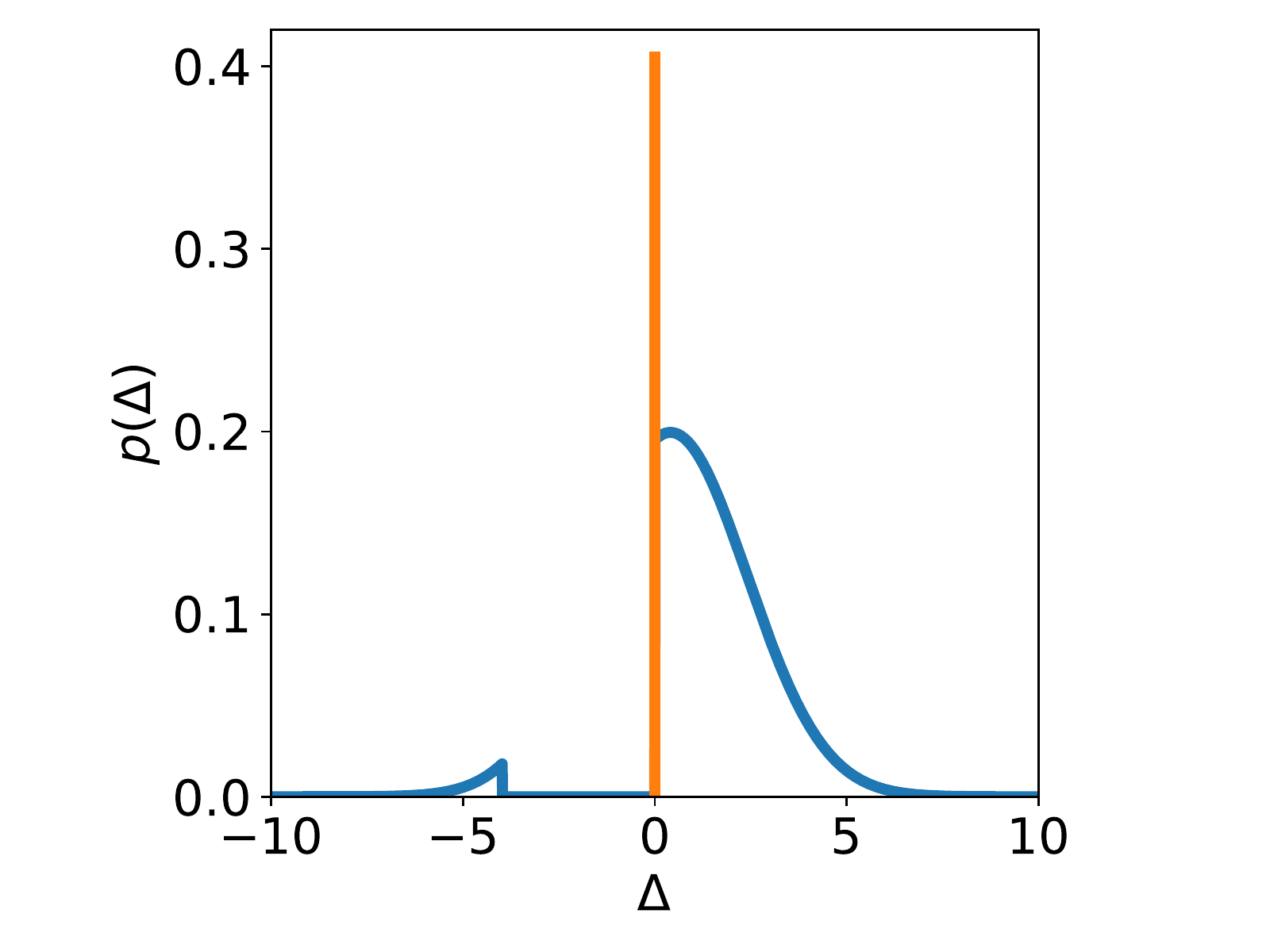}
	\hspace{0.05\linewidth}
	\includegraphics[width=0.41\linewidth]{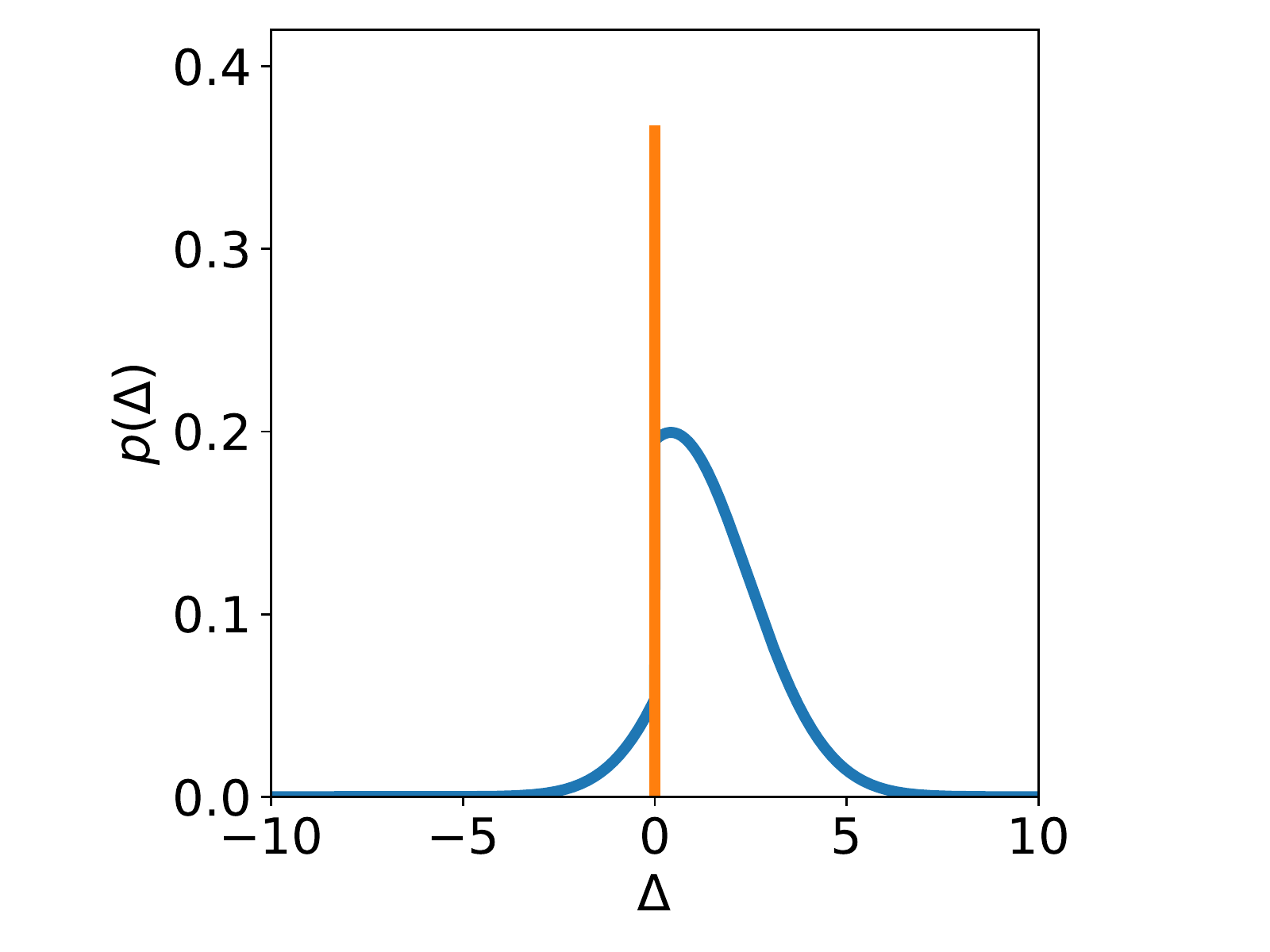}
	\caption{Left:  Distribution of violation strengths $\Delta$ for the step function (left) and the ramp function (right). The height of the line at $\Delta=0$ gives the prefactor of the respective $\delta$ function. Parameters: $A=B=\gamma=1$, $ \sigma=2, \alpha=3$. 	
	}
	\label{Fig: stability distributions}
\end{figure*}

For the ramp cost function (\ref{deframp}) the calculation is similar. Eq.~(\ref{Equ: Minimal Delta step}) is modified to 
\begin{equation}\label{Equ: Minimal Delta ramp}
\Delta_0(t)=
\begin{cases}
\sigma(\kappa-t) & \text{if } t<\kappa \\ \nonumber
0 & \text{if } \kappa \leq t \leq \kappa+\sigma x \\
\sigma(x\sigma+\kappa-t) & \text{if } t > \kappa+\sigma x
\end{cases}       
\end{equation}
and $G_E$ acquires the form 
\begin{align}
\frac{G_E}{\beta}=-\frac{1}{2x}\int_{\kappa}^{\kappa+x\sigma}&\!\!\!\!\!\! Dt\, (\kappa-t)^2 
      +\sigma \left(\kappa+\frac{x \sigma}{2}\right) H(\kappa+x\sigma)\nonumber \\
    &-\frac{\sigma }{\sqrt{2\pi}}\exp\left\{-\frac{(\kappa+x\sigma)^2}{2}\right\}.
\end{align}

The typical minimal cost per equation becomes
\begin{align}\nonumber
\langle E_\mathrm{min}\rangle
   &=\text{extr}_{x,\kappa} \nonumber \Bigg\{\!\!\!-\frac{1}{2\alpha x}+\frac{\lambda^2\kappa^2}{2\alpha x}
            -\! \sigma\left(\! \kappa \!+\!\frac{x \sigma}{2}\right) H(\kappa \!+\!x\sigma) \\
+\frac{1}{2x}& \int_{\kappa}^{\kappa+x\sigma} \!\!\!\! Dt (\kappa-t)^2
+\frac{\sigma }{\sqrt{2\pi}}\exp\left\{-\frac{(\kappa+x\sigma)^2}{2}\right\}\Bigg\},
\end{align}
complemented by the saddle point equations
\begin{align}
1-\lambda^2\kappa^2=& \,\alpha\sigma^2 x^2 H(\kappa+\sigma x)+\alpha \int_{\kappa}^{\kappa+x\sigma}\!\!\!\!\!Dt(\kappa-t)^2, \label{Equ: Saddle point perceptron 1}
\\
\lambda^2 \kappa=& \, \alpha x \sigma H(k+\sigma x)-\alpha  \int_{\kappa}^{\kappa+x\sigma}\!\!\!\!\!Dt (\kappa-t).\label{Equ: Saddle point perceptron 2}
\end{align}

The final result for the typical minimal cost per equation assumes the form 
\begin{align}\label{Equ: Vmin Perceptron}
\left<E_{\min}\right>=-\sigma (\kappa+x \sigma) H(\kappa+x \sigma)+\frac{\sigma}{\sqrt{2 \pi}} e^{-(\kappa+x \sigma)^2/2},
\end{align}
where again the values of the saddle-point variables $\kappa$ and $x$ as determined from the numerical solution of~(\ref{Equ: Saddle point perceptron 1}) and (\ref{Equ: Saddle point perceptron 2}) have to be plugged into this expression. 

In the right part of Figure~\ref{Fig: Minimal energies} analytical and numerical results for $\left< E_{\min} \right>$ for the ramp cost function are compared. In this case, it is straightforward to get numerical results since the minimal value of $v(\mathbf{y})$ can be determined using standard tools as the scipy.optimize.minimize function in Python. Very good agreement between analytical and numerical results is found. 

The results on the minimal cost per equation shown in Fig.~\ref{Fig: Minimal energies} can be specified in more detail by determining the distribution $p_{\min}(\Delta)$ of the "violation strengths" $\Delta_\mu$ defined in \eqref{defDelta} in the limit $\beta\to\infty$. In \cite{griniasty1991learning} it is shown that this distribution is given by
\begin{equation}\label{Equ: Distribution of Delta}
p_{\min}(\Delta)=\int \!\! Dt \, \delta(\Delta-\Delta_0(t)).
\end{equation}
In this way we find for the step cost function
\begin{align}
p_{\min}(\Delta)=&\delta(\Delta) \int_{\kappa}^{\kappa+ \sqrt{2x}}\!\!\!\! Dt\nonumber
           \\\label{Equ: Distribution Derrida}
   +\Big(\Theta(\Delta)&+\Theta(-\Delta-\sigma\sqrt{2 x})\Big) 
     \frac{1}{\sqrt{2 \pi}\sigma} e^{-(\kappa-\frac{\Delta}{\sigma})^2/2} 
\end{align}
and for the ramp function 
\begin{align}
p_{\min}(\Delta)=&\delta(\Delta)\,\int_{\kappa}^{\kappa+\sigma x}\!\!\!\! Dt 
   + \Theta(\Delta) \frac{1}{\sqrt{2 \pi}\sigma}e^{-(\kappa-\frac{\Delta}{\sigma})^2/2} \nonumber
	\\\label{Equ: Distribution of Delta perceptron}
	&+\Theta(-\Delta)\frac{1}{\sqrt{2 \pi}\sigma}e^{-(-\frac{\Delta}{\sigma}+\sigma x +\kappa)^2/2}. 
\end{align}

Figure \ref{Fig: stability distributions} compares these two distributions in the interesting regime $\alpha > \alpha_c$. The left part shows $p_{\min}$ for the step function. The $\delta$-peak at $\Delta=0$ is caused by constraints that are fulfilled as equalities whereas the part of the Gaussian for $\Delta>0$ represents constraints which are more than just barely fulfilled. Note the gap of width $\sigma \sqrt{2 x}$ in the negative part of the distribution. This gap is due to the fact that the step cost function does not differentiate between heavily violated constraints and those only slightly violated. It therefore prefers few ``gross mistakes'' to many tiny ones. In this way it selects the row vectors $\mathbf{a}_\mu$ that are ``most crucial'' for the cone to contain the vector $\mathbf{b}$ as discussed at the beginning of this section. In contrast, the distribution $p_{\min}(\Delta)$ for the ramp cost function is gapless, see the right part of Figure~\ref{Fig: stability distributions}. Since the ramp function punishes strong violations of constraints more severely than slight violations its optimal value is realized with many but small violations. It therefore characterizes some average distance of $\mathbf{b}$ from the boundary of the cone. No gap in the distribution is therefore to be expected. 
Note that the typical minimal cost per equation is recovered from the distributions $p_{\min}(\Delta)$ by:
\begin{align}
 \left<E_{\min} \right>	= \int d\Delta \,E(\Delta) \, p_{\min} (\Delta) .
\end{align}

\section{Conclusion} \label{Sec: Conclusion}
Large systems of random linear equations exhibit intriguing properties if their variables are constrained to be non-negative. These systems show a sharp transition from a phase where a solution can be found with probability one to a phase where typically no such solution exists. The transition line depends only on the statistical properties of the systems and is hence self-averaging. 

In the present paper we showed that the mapping used in \cite{landmann2019non} to determine the critical line is also the clue to characterize the two phases away from criticality. To this end we introduced suitable quantities describing the phases and determined their typical values as function of the statistical parameters of the random ensembles and the ratio between the number of variables and the number of equations. The unsolvable phase can be characterized by the typical residual norm of the system of equations. It measures how 'far away' the system is from being solvable. In the solvable phase two measures for the robustness of solutions were introduced and analyzed. 

Our analytical investigations became possible by first mapping the problem onto a dual one using Farkas' lemma, and then analyzing this dual problem with the help of the replica method. Technically, our results depend on the correctness of the assumption of replica symmetry. In the unsolvable phase, $\alpha\leq\alpha_c$, the solution space of the dual problem is connected and it is reasonable to expect replica symmetry to hold. On the other hand, the solvable phase is characterized by a disconnected solution space of the dual problem and replica symmetry breaking may come into play \cite{GaDe,bouten1994replica,griniasty1991learning}.

\begin{acknowledgments}
 We would like to thank Mattes Heerwagen and Sebastian Rosmej for fruitful discussions and Imre Kondor for interesting correspondence. Financial support from the German Science Foundation under project EN 278/10-1 is gratefully acknowledged.
\end{acknowledgments}

\bibliography{bibliography.bib}

\begin{thebibliography}{22}%
\makeatletter
\providecommand \@ifxundefined [1]{%
 \@ifx{#1\undefined}
}%
\providecommand \@ifnum [1]{%
 \ifnum #1\expandafter \@firstoftwo
 \else \expandafter \@secondoftwo
 \fi
}%
\providecommand \@ifx [1]{%
 \ifx #1\expandafter \@firstoftwo
 \else \expandafter \@secondoftwo
 \fi
}%
\providecommand \natexlab [1]{#1}%
\providecommand \enquote  [1]{``#1''}%
\providecommand \bibnamefont  [1]{#1}%
\providecommand \bibfnamefont [1]{#1}%
\providecommand \citenamefont [1]{#1}%
\providecommand \href@noop [0]{\@secondoftwo}%
\providecommand \href [0]{\begingroup \@sanitize@url \@href}%
\providecommand \@href[1]{\@@startlink{#1}\@@href}%
\providecommand \@@href[1]{\endgroup#1\@@endlink}%
\providecommand \@sanitize@url [0]{\catcode `\\12\catcode `\$12\catcode
  `\&12\catcode `\#12\catcode `\^12\catcode `\_12\catcode `\%12\relax}%
\providecommand \@@startlink[1]{}%
\providecommand \@@endlink[0]{}%
\providecommand \url  [0]{\begingroup\@sanitize@url \@url }%
\providecommand \@url [1]{\endgroup\@href {#1}{\urlprefix }}%
\providecommand \urlprefix  [0]{URL }%
\providecommand \Eprint [0]{\href }%
\providecommand \doibase [0]{http://dx.doi.org/}%
\providecommand \selectlanguage [0]{\@gobble}%
\providecommand \bibinfo  [0]{\@secondoftwo}%
\providecommand \bibfield  [0]{\@secondoftwo}%
\providecommand \translation [1]{[#1]}%
\providecommand \BibitemOpen [0]{}%
\providecommand \bibitemStop [0]{}%
\providecommand \bibitemNoStop [0]{.\EOS\space}%
\providecommand \EOS [0]{\spacefactor3000\relax}%
\providecommand \BibitemShut  [1]{\csname bibitem#1\endcsname}%
\let\auto@bib@innerbib\@empty
\bibitem [{\citenamefont {Edwards}\ and\ \citenamefont
  {Anderson}(1975)}]{edwards1975theory}%
  \BibitemOpen
  \bibfield  {author} {\bibinfo {author} {\bibfnamefont {S.~F.}\ \bibnamefont
  {Edwards}}\ and\ \bibinfo {author} {\bibfnamefont {P.~W.}\ \bibnamefont
  {Anderson}},\ }\href@noop {} {\bibfield  {journal} {\bibinfo  {journal}
  {Journal of Physics F: Metal Physics}\ }\textbf {\bibinfo {volume} {5}},\
  \bibinfo {pages} {965} (\bibinfo {year} {1975})}\BibitemShut {NoStop}%
\bibitem [{\citenamefont {Sherrington}\ and\ \citenamefont
  {Kirkpatrick}(1975)}]{sherrington1975solvable}%
  \BibitemOpen
  \bibfield  {author} {\bibinfo {author} {\bibfnamefont {D.}~\bibnamefont
  {Sherrington}}\ and\ \bibinfo {author} {\bibfnamefont {S.}~\bibnamefont
  {Kirkpatrick}},\ }\href@noop {} {\bibfield  {journal} {\bibinfo  {journal}
  {Physical Review Letters}\ }\textbf {\bibinfo {volume} {35}},\ \bibinfo
  {pages} {1792} (\bibinfo {year} {1975})}\BibitemShut {NoStop}%
\bibitem [{\citenamefont {Monasson}\ \emph {et~al.}(1999)\citenamefont
  {Monasson}, \citenamefont {Zecchina}, \citenamefont {Kirkpatrick},
  \citenamefont {Selman},\ and\ \citenamefont
  {Troyansky}}]{monasson1999determining}%
  \BibitemOpen
  \bibfield  {author} {\bibinfo {author} {\bibfnamefont {R.}~\bibnamefont
  {Monasson}}, \bibinfo {author} {\bibfnamefont {R.}~\bibnamefont {Zecchina}},
  \bibinfo {author} {\bibfnamefont {S.}~\bibnamefont {Kirkpatrick}}, \bibinfo
  {author} {\bibfnamefont {B.}~\bibnamefont {Selman}}, \ and\ \bibinfo {author}
  {\bibfnamefont {L.}~\bibnamefont {Troyansky}},\ }\href@noop {} {\bibfield
  {journal} {\bibinfo  {journal} {Nature}\ }\textbf {\bibinfo {volume} {400}},\
  \bibinfo {pages} {133} (\bibinfo {year} {1999})}\BibitemShut {NoStop}%
\bibitem [{\citenamefont {Mezard}\ and\ \citenamefont
  {Montanari}(2009)}]{mezard2009information}%
  \BibitemOpen
  \bibfield  {author} {\bibinfo {author} {\bibfnamefont {M.}~\bibnamefont
  {Mezard}}\ and\ \bibinfo {author} {\bibfnamefont {A.}~\bibnamefont
  {Montanari}},\ }\href@noop {} {\emph {\bibinfo {title} {{Information,
  Physics, and Computation}}}}\ (\bibinfo  {publisher} {Oxford University
  Press},\ \bibinfo {year} {2009})\BibitemShut {NoStop}%
\bibitem [{\citenamefont {Engel}\ and\ \citenamefont {Van~den
  Broeck}(2001)}]{engel2001statistical}%
  \BibitemOpen
  \bibfield  {author} {\bibinfo {author} {\bibfnamefont {A.}~\bibnamefont
  {Engel}}\ and\ \bibinfo {author} {\bibfnamefont {C.}~\bibnamefont {Van~den
  Broeck}},\ }\href@noop {} {\emph {\bibinfo {title} {Statistical mechanics of
  learning}}}\ (\bibinfo  {publisher} {Cambridge University Press},\ \bibinfo
  {year} {2001})\BibitemShut {NoStop}%
\bibitem [{\citenamefont {Feng}\ and\ \citenamefont
  {Zhang}(2007)}]{feng2007rank}%
  \BibitemOpen
  \bibfield  {author} {\bibinfo {author} {\bibfnamefont {X.}~\bibnamefont
  {Feng}}\ and\ \bibinfo {author} {\bibfnamefont {Z.}~\bibnamefont {Zhang}},\
  }\href@noop {} {\bibfield  {journal} {\bibinfo  {journal} {Applied
  Mathematics and Computation}\ }\textbf {\bibinfo {volume} {185}},\ \bibinfo
  {pages} {689} (\bibinfo {year} {2007})}\BibitemShut {NoStop}%
\bibitem [{\citenamefont {May}(1972)}]{may1972will}%
  \BibitemOpen
  \bibfield  {author} {\bibinfo {author} {\bibfnamefont {R.~M.}\ \bibnamefont
  {May}},\ }\href@noop {} {\bibfield  {journal} {\bibinfo  {journal} {Nature}\
  }\textbf {\bibinfo {volume} {238}},\ \bibinfo {pages} {413} (\bibinfo {year}
  {1972})}\BibitemShut {NoStop}%
\bibitem [{\citenamefont {Eigen}\ and\ \citenamefont
  {Schuster}(1978)}]{Schuster}%
  \BibitemOpen
  \bibfield  {author} {\bibinfo {author} {\bibfnamefont {M.}~\bibnamefont
  {Eigen}}\ and\ \bibinfo {author} {\bibfnamefont {P.}~\bibnamefont
  {Schuster}},\ }\href@noop {} {\bibfield  {journal} {\bibinfo  {journal}
  {Naturwissenschaften}\ }\textbf {\bibinfo {volume} {65}},\ \bibinfo {pages}
  {7} (\bibinfo {year} {1978})}\BibitemShut {NoStop}%
\bibitem [{\citenamefont {Tikhonov}\ and\ \citenamefont
  {Monasson}(2017)}]{tikhonov2017collective}%
  \BibitemOpen
  \bibfield  {author} {\bibinfo {author} {\bibfnamefont {M.}~\bibnamefont
  {Tikhonov}}\ and\ \bibinfo {author} {\bibfnamefont {R.}~\bibnamefont
  {Monasson}},\ }\href@noop {} {\bibfield  {journal} {\bibinfo  {journal}
  {Physical Review Letters}\ }\textbf {\bibinfo {volume} {118}},\ \bibinfo
  {pages} {048103} (\bibinfo {year} {2017})}\BibitemShut {NoStop}%
\bibitem [{\citenamefont {Polettini}\ and\ \citenamefont
  {Esposito}(2014)}]{PoEs}%
  \BibitemOpen
  \bibfield  {author} {\bibinfo {author} {\bibfnamefont {M.}~\bibnamefont
  {Polettini}}\ and\ \bibinfo {author} {\bibfnamefont {M.}~\bibnamefont
  {Esposito}},\ }\href@noop {} {\bibfield  {journal} {\bibinfo  {journal} {The
  Journal of chemical physics}\ }\textbf {\bibinfo {volume} {141}},\ \bibinfo
  {pages} {024117} (\bibinfo {year} {2014})}\BibitemShut {NoStop}%
\bibitem [{\citenamefont {Schnakenberg}(1979)}]{Schnakenberg}%
  \BibitemOpen
  \bibfield  {author} {\bibinfo {author} {\bibfnamefont {J.}~\bibnamefont
  {Schnakenberg}},\ }\href@noop {} {\bibfield  {journal} {\bibinfo  {journal}
  {Journal of Theoretical Biology}\ }\textbf {\bibinfo {volume} {81}},\
  \bibinfo {pages} {389} (\bibinfo {year} {1979})}\BibitemShut {NoStop}%
\bibitem [{\citenamefont {Kondor}\ \emph {et~al.}(2017)\citenamefont {Kondor},
  \citenamefont {Papp},\ and\ \citenamefont {Caccioli}}]{kondor2017analytic}%
  \BibitemOpen
  \bibfield  {author} {\bibinfo {author} {\bibfnamefont {I.}~\bibnamefont
  {Kondor}}, \bibinfo {author} {\bibfnamefont {G.}~\bibnamefont {Papp}}, \ and\
  \bibinfo {author} {\bibfnamefont {F.}~\bibnamefont {Caccioli}},\ }\href@noop
  {} {\bibfield  {journal} {\bibinfo  {journal} {Journal of Statistical
  Mechanics: Theory and Experiment}\ }\textbf {\bibinfo {volume} {2017}},\
  \bibinfo {pages} {123402} (\bibinfo {year} {2017})}\BibitemShut {NoStop}%
\bibitem [{\citenamefont {Berg}\ and\ \citenamefont
  {Engel}(1998)}]{berg1998matrix}%
  \BibitemOpen
  \bibfield  {author} {\bibinfo {author} {\bibfnamefont {J.}~\bibnamefont
  {Berg}}\ and\ \bibinfo {author} {\bibfnamefont {A.}~\bibnamefont {Engel}},\
  }\href@noop {} {\bibfield  {journal} {\bibinfo  {journal} {Physical Review
  Letters}\ }\textbf {\bibinfo {volume} {81}},\ \bibinfo {pages} {4999}
  (\bibinfo {year} {1998})}\BibitemShut {NoStop}%
\bibitem [{\citenamefont {Landmann}\ and\ \citenamefont
  {Engel}(ress)}]{landmann2019non}%
  \BibitemOpen
  \bibfield  {author} {\bibinfo {author} {\bibfnamefont {S.}~\bibnamefont
  {Landmann}}\ and\ \bibinfo {author} {\bibfnamefont {A.}~\bibnamefont
  {Engel}},\ }\href@noop {} {\bibfield  {journal} {\bibinfo  {journal} {Physica
  A: Statistical Mechanics and its Applications}\ ,\ \bibinfo {pages} {122544}}
  (\bibinfo {year} {In Press})}\BibitemShut {NoStop}%
\bibitem [{\citenamefont {Farkas}(1902)}]{farkas1902theorie}%
  \BibitemOpen
  \bibfield  {author} {\bibinfo {author} {\bibfnamefont {J.}~\bibnamefont
  {Farkas}},\ }\href@noop {} {\bibfield  {journal} {\bibinfo  {journal}
  {Journal f{\"u}r die reine und angewandte Mathematik}\ }\textbf {\bibinfo
  {volume} {124}},\ \bibinfo {pages} {1} (\bibinfo {year} {1902})}\BibitemShut
  {NoStop}%
\bibitem [{\citenamefont {M{\'e}zard}\ \emph {et~al.}(1987)\citenamefont
  {M{\'e}zard}, \citenamefont {Parisi},\ and\ \citenamefont
  {Virasoro}}]{mezard1987spin}%
  \BibitemOpen
  \bibfield  {author} {\bibinfo {author} {\bibfnamefont {M.}~\bibnamefont
  {M{\'e}zard}}, \bibinfo {author} {\bibfnamefont {G.}~\bibnamefont {Parisi}},
  \ and\ \bibinfo {author} {\bibfnamefont {M.}~\bibnamefont {Virasoro}},\
  }\href@noop {} {\emph {\bibinfo {title} {Spin glass theory and beyond}}}\
  (\bibinfo  {publisher} {World Scientific Publishing Company},\ \bibinfo
  {year} {1987})\BibitemShut {NoStop}%
\bibitem [{Note1()}]{Note1}%
  \BibitemOpen
  \bibinfo {note} {We used the non-negative least squares solver nnls from the
  scipy.optimize package in Python.}\BibitemShut {Stop}%
\bibitem [{\citenamefont {Engel}\ and\ \citenamefont {Van~den
  Broeck}(1993)}]{engel1993systems}%
  \BibitemOpen
  \bibfield  {author} {\bibinfo {author} {\bibfnamefont {A.}~\bibnamefont
  {Engel}}\ and\ \bibinfo {author} {\bibfnamefont {C.}~\bibnamefont {Van~den
  Broeck}},\ }\href@noop {} {\bibfield  {journal} {\bibinfo  {journal}
  {Physical Review Letters}\ }\textbf {\bibinfo {volume} {71}},\ \bibinfo
  {pages} {1772} (\bibinfo {year} {1993})}\BibitemShut {NoStop}%
\bibitem [{Note2()}]{Note2}%
  \BibitemOpen
  \bibinfo {note} {Both functions have an equivalent in the theory of neural
  networks: The step function corresponds to the Gardner-Derrida cost function
  \cite {GaDe}, the ramp function to the perceptron cost function \cite
  {griniasty1991learning}.}\BibitemShut {Stop}%
\bibitem [{\citenamefont {Gardner}\ and\ \citenamefont {Derrida}(1988)}]{GaDe}%
  \BibitemOpen
  \bibfield  {author} {\bibinfo {author} {\bibfnamefont {E.}~\bibnamefont
  {Gardner}}\ and\ \bibinfo {author} {\bibfnamefont {B.}~\bibnamefont
  {Derrida}},\ }\href@noop {} {\bibfield  {journal} {\bibinfo  {journal}
  {Journal of Physics A: Mathematical and general}\ }\textbf {\bibinfo {volume}
  {21}},\ \bibinfo {pages} {271} (\bibinfo {year} {1988})}\BibitemShut
  {NoStop}%
\bibitem [{\citenamefont {Griniasty}\ and\ \citenamefont
  {Gutfreund}(1991)}]{griniasty1991learning}%
  \BibitemOpen
  \bibfield  {author} {\bibinfo {author} {\bibfnamefont {M.}~\bibnamefont
  {Griniasty}}\ and\ \bibinfo {author} {\bibfnamefont {H.}~\bibnamefont
  {Gutfreund}},\ }\href@noop {} {\bibfield  {journal} {\bibinfo  {journal}
  {Journal of Physics A: Mathematical and General}\ }\textbf {\bibinfo {volume}
  {24}},\ \bibinfo {pages} {715} (\bibinfo {year} {1991})}\BibitemShut
  {NoStop}%
\bibitem [{\citenamefont {Bouten}\ and\ \citenamefont
  {Derrida}(1994)}]{bouten1994replica}%
  \BibitemOpen
  \bibfield  {author} {\bibinfo {author} {\bibfnamefont {M.}~\bibnamefont
  {Bouten}}\ and\ \bibinfo {author} {\bibfnamefont {B.}~\bibnamefont
  {Derrida}},\ }\href@noop {} {\bibfield  {journal} {\bibinfo  {journal}
  {Journal of Physics A: Mathematical, Nuclear and General}\ }\textbf {\bibinfo
  {volume} {27}},\ \bibinfo {pages} {6021} (\bibinfo {year}
  {1994})}\BibitemShut {NoStop}%
\end{thebibliography}%

\end{document}